# Forecasting mortality using Google trend

Fu-Chun Yeh

*Department of Aviation Electronics, China University of Science and Technology, Taiwan*

**Abstract**

In this paper, the motility model for the developed country, which United State possesses the largest economy in the world and thus serves as an ideal representation, is investigated. Early surveillance of the causes of death is critical which can allow the preparation of preventive steps against critical disease such as dengue fever. Studies reported that some search queries, especially those diseases related terms on Google Trends are essential. To this end, we include either main cause of death or the extended or the more general terminologies from Google Trends to decode the mortality related terms using the Wiener Cascade Model. Using time series and Wavelet scalogram of search terms, the patterns of search queries are categorized into different levels of periodicity. The results include (1) the decoding trend, (2) the features importance, and (3) the accuracy of the decoding patterns. Three scenarios regard predictors include the use of (1) all 19 features, (2) the top ten most periodic predictors, or (3) the ten predictors with highest weighting. All search queries spans from December 2013 – December 2018. The results show that search terms with both higher weight and annual periodic pattern contribute more in forecasting the word "die"; however, only predictors with higher weight are valuable to forecast the word "death".

## 1. Introduction

Big data, a large volume of data increases on a day-to-day basis, may reveal abundant information of various activities providing precious databases to discover fundamental regulation or behavior regard the hidden complicated activities of our daily life [1–4]. Quantitative prediction of mortality trend is essential for government and healthy agency. The list of the causes of human deaths worldwide exerts immense impacts on governmental policy for medical affairs, as well as personal disease prevention, requiring sufficient attention to this topic.

Traditionally, mortality trend was estimated by the death and medical records, which require gathering information and immense statistical investigations, and thus are less able to provide timely information and prediction. Later, internet allowed monitoring transmission of diseases via trace reports. Data log from Wikipedia already applied to predict transmission of diseases in several countries [5,6]. However, the text data, which fails to reflect scales smaller than country, limits the investigation regard detailed location [6]. Nowadays, online sources which consists immense information are available.

Several reports applied information from search queries on Google, such as Google Dengue Trends and Google Flu Trends, which terminated the service in 2015, in predicting and monitoring influenza pandemic [7–9]. The Google search engine allows the access of dynamical changes of search terms on their search volume, which is known as Google Trends and assessable for public. By using search queries from Google Trends, previous reports showed successful forecasting of influenza and dengue fever in South Korea in Indonesia [10,11]. Unlike data log from Wikipedia, Google Trends allow the fine request by geographical regions and temporal duration. United State, as the world largest economy, raises great impact over its death causes. Forecasting the impacts from various mortal factors is essential for planning an appropriate response strategy [5]. In this paper, we investigate the availability of applying search terms of critical causes of death from Google Trends to predict mortality dynamics, as well as optimize features require to assure promising decoding, which could provide information of rhythmic patterns of all the critical diseases and the associated search terms as well.

In the present study, we investigate search terms for time period between 2013 and 2018, and expect that these query search data not only present the historical occurrence of the diseases, but may contribute to the decoding of the future mortality trends as well. Our results support our hypothesis that search terms of critical death of cause can predict death. Therefore, we may search for more information about the death dynamics, and take precautions against risky factors. Our findings suggest that the search query volumes from Google Trends for some search terms, either periodic or non-periodic trending data, may contribute to the determination of strategies for medical affairs.

## 2. Materials and methods

*2.1. Data*

Table 1 summarizes 19 search indexes as candidate of predictors, and two other search indexes as dependent variables. The search queries of terms which include "die" and "death" were set as the dependent variables. 19 Google Trend search index of diseases and death causes may serve as predictors of all models. The present study focus on the search queries from United State (http://www.google.com/trend). All data spanned from December 2013 to December 2018. All data were sampled by weeks. Google Trend search queries related to mortality is determined based on related main death causes (i.e., "AIDS", "Alzheimer", "Breast Cancer", "Cirrhosis", "Diabetes", "Diarrhoeal", "Heart Disease", "Kidney Cancer", "Lung Cancer", "Malaria", "Obstructive Pulmonary Disease", "Respiration Infection", "Stomach Cancer", "Stroke", and "Tuberculosis") according to World Health Organization, and the more general or extended search terms (i.e., "Cancer", "Car Accident", "Flu", and "Sick").

**Table 1.** Search terms, role as predictors or variables, existence of periodic patterns, and weighting of all search terms.

| No. | Search Term | Role | Periodic | Weight (die/death) |
|---|---|---|---|---|
| 1 | **AIDS** | Predictor | 6 months | 12/11 |
| 2 | **Alzheimer** | Predictor | 6 months | 9/14 |
| 3 | **Breast Cancer** | Predictor | 1 year | 13/16 |
| 4 | **Cancer** | Predictor | 1 year | 1/5 |
| 5 | **Car Accident** | Predictor | N.A. | 7/10 |
| 6 | **Cirrhosis** | Predictor | 6 months | 8/6 |
| A | **Death** | Variable | 6 months | |
| 7 | **Diabetes** | Predictor | 6 months/1 year | 3/3 |
| 8 | **Diarrhoeal** | Predictor | N.A. | 18/18 |
| B | **Die** | Variable | 1 year | |
| 9 | **Flu** | Predictor | 1 year | 17/13 |
| 10 | **Heart Disease** | Predictor | 1 year | 4/4 |
| 11 | **Kidney Cancer** | Predictor | 1 year | 6/7 |
| 12 | **Lung Cancer** | Predictor | 6 months | 2/2 |
| 13 | **Malaria** | Predictor | 6 months/1 year | 16/17 |
| 14 | **Obstructive Pulmonary Disease** | Predictor | 6 months/1 year | 14/15 |
| 15 | **Respiratory Infection** | Predictor | 1 year | 15/12 |
| 16 | **Sick** | Predictor | 1 year | 10/1 |
| 17 | **Stomach Cancer** | Predictor | N.A. | 11/9 |
| 18 | **Stroke** | Predictor | 1 year | 5/8 |
| 19 | **Tuberculosis** | Predictor | 6 months | 19/19 |

*2.2 Wiener Cascade Model*

The search terms "die" or "death" from several potential causes of death from Google Trends are decoded using Wiener Cascade Model. The main causes of death and the extended or more general terminologies include 19 search terms (see Table 1 for details). The sampling rate is 52 count/year. Weekly search query of each term is represented as time series and its time-frequency representation (Fig. 1). We calculate spectral power using the Morlet Wavelet transform. The frequency domain ranges between 0.5 count/year (2 years) and 4 count/year (3 months). We investigate the selections of predictive terms into three types of criterion: (1) Use all 19 potential search terms as features. (2) Investigate time and frequency representations of all search terms, and rank their tendency as cyclic patterns (see Fig. 1C as an example), then select the top 10 features. Of note the feature selection was performed on predictors, not "die" (Fig. 1A) or "death" (Fig. 1B). (3) Select the top 10 high-weighted features (Fig. 3) according to the performances of using all 19 potential search terms as predictors. The Wiener Cascade Model [12], as the 3rd order polynomial, convolve the output of a linear multi-input and single-output filter with a static nonlinearity [13–15] as the iterative adjustment of the output, and decode search terms "die" or "death" as dependent variable. The Wiener filter is formulated as

$$X(t) = \sum_{j=-52}^{0} \sum_{i=1}^{n} A_{ij} x(i, t+j),$$

where $X(t)$, $x(i,j)$, $A_{ij}$, $n$, and $t$ represent the search terms "die" or "death", the value of feature $i$ at time lag $j$, the filter coefficient for the time lag, the number of features, as well as the current time, respectively. A time lags of 52 weeks are included as stated in the equation. The coefficients can be estimated by the ridge regression $a = (f^T f + \lambda I)^{-1} f^T F$, where $\lambda$ and $I$ denote the regularization constant and the identity matrix, respectively.

The accuracy of the decoding is measured by calculating the spearman correlation (rho) between predicted and actual mortality associated terminologies (i.e., "die" or "death"), as well as the performance of its significant test [15], which implies whether the accuracy of predicted mortality levels correlate with continuous search queries of the dependent variables. On the other side, the mean square error (MSE) is introduced to evaluate the accuracy of the decoding as well; i.e., we tested whether the variation of the estimation was confined within a small range. We use 5-fold cross-validation. That is, 4 of the 5 folds are served as training datasets to decode the other one fold. The above procedure is repeated for all three selections of predictors, and the measures of the accuracy of the decoding are summarized and compared in Table 2.

## 3. Results and discussions

We include 19 potential search terms to forecast two other mortality related search terms (Table 1). Each dataset performs 5 years of this weekly search query. Examples of time series and Wavelet scalogram of search terms "die" and "death" are presented in Fig. 1A and 1B, respectively. Of note the former one (Fig. 1A) is more annual periodic (i.e., approximately one cycle per year), whereas the latter one possess less cyclic pattern (Fig. 1B). Fig. 1C demonstrates a typical example of predictor (e.g., respiration infection) with simple annual cyclic pattern, whereas Fig. 1D presents an example of predictor (e.g., Malaria) with both annual and semiannual periodic patterns.

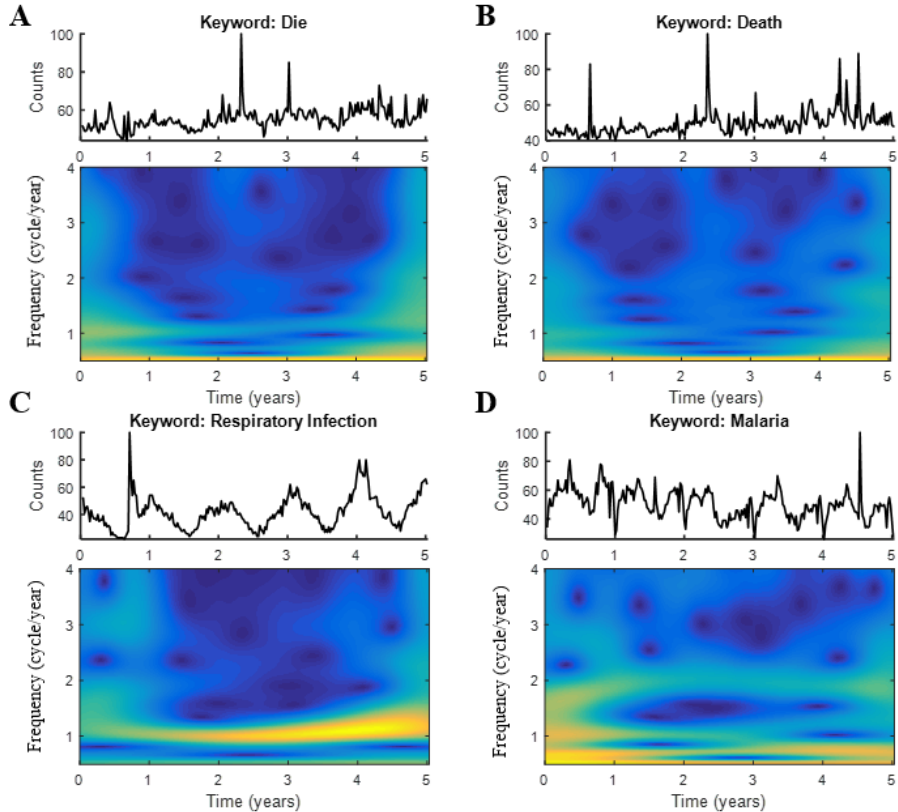

**Fig. 1.** Time series and Wavelet scalogram of search terms (A) "Die", (B) "Death", (C) "Respiratory Infection", and (D) "Malaria".

Fig. 2 presents the decoded (red tracks) and actual (black tracks) trends of search terms "die" or "death". Fig. 2A shows the performance of using all 19 search terms as predictors to decode the search term "die" (rho = 0.49, p < 0.001; MSE = 23.24), which is superior in accuracy of decoding than that of the search term "death" as shown in Fig. 2B (rho = 0.25, p < 0.001; MSE = 52.61). If considering only the top ten most annual periodic search terms, the accuracy in decoding the search term "die" (Fig. 2C) is quite close (rho = 0.44), whereas the performance for the search term "death" (Fig. 2D) fail to obtain significant correlation between the decoded (red tracks) and actual (black tracks) trends (rho = 0.07, p = 0.32; MSE = 78.00). Table 1 also summarizes the observation of periodic patterns of all 21 search terms. The top ten most periodic predictors include "Breast Cancer", "Cancer", "Diabetes", "Flu", "Heart Disease", "Kidney", "Malaria" (Fig. 1D), "Respiration Infection" (Fig. 1C), "Sick", and "Stroke".

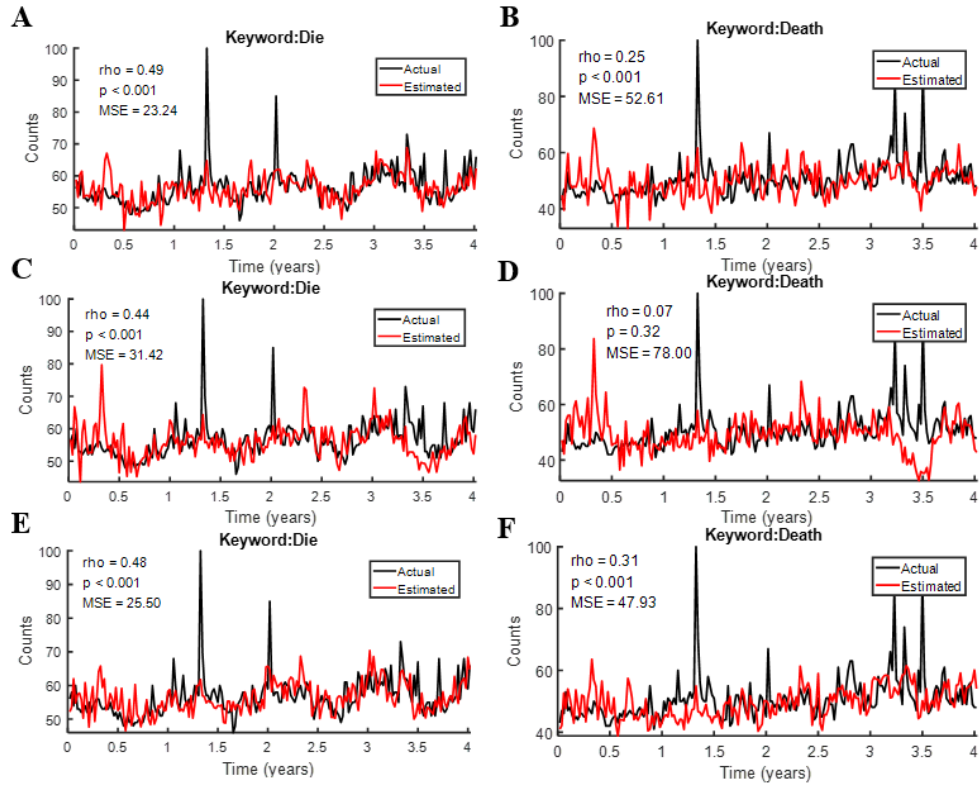

**Fig. 2.** The decoded (red tracks) and actual (black tracks) trends of search terms "die" or "death". Performances of using three different types of selections of search terms are applied. (A) and (B) present the decoding results of "die" and "death" using all 19 predictors. (C) and (D) show the performance of "die" and "death" using only the top ten most periodic predictors. (E) and (F) demonstrate the trends of "die" and "death" using only the ten predictors with highest weighting in (A) and (B).

The color plots in Fig. 3 present the contribution of features of each selected search query and at each time lag for mortality forecasting across all five-year data. To visualize the weight distribution across features intuitively, the sum of weight for selected features are provided as the bar plot in Fig. 3. Table 1 also summarizes the rank of all the 19 features importance according to the bar plot in Fig. 3A (i.e., "die") and 3B (i.e., "death"). The ten highest weighting predictors include "Cancer", "Car Accident", "Cirrhosis", "Diabetes", "Heart Disease", "Kidney", "Lung Cancer", "Sick", "Stomach Cancer", and "Stroke". When including only these ten search terms with the highest weight, the accuracy in decoding for the search term "die" (Fig. 2E) is quite close (rho = 0.48, p < 0.001; MSE = 25.50) to that with all 19 predictors (Fig. 2A; rho = 0.49, p < 0.001; MSE = 23.24), whereas the performance for the search term "death" (Fig. 2F; rho = 0.31, p < 0.001; MSE = 47.93) even reveal superior accuracy than that with all 19 predictors (Fig. 2F; rho = 0.25, p < 0.001; MSE = 52.61). If using only the top ten most periodic predictors, the sum of weight for features show the three search queries with highest weights to forecast "die" are "Cancer", "Diabetes", and "Heart Disease", respectively

(Fig. 3C). As for the case to decode "death", the three search queries are "Cancer", "Sick" and "Diabetes", respectively (Fig. 3D); of note this prediction is not significant as shown in Fig. 2D. The lowest panels in Fig. 3 present the features importance for mortality forecasting using only the ten predictors with highest weighting according to the situations with all 19 predictors included (i.e., Fig. 3A and 3B). The three most prominent search queries to predict "die" are "Cancer", "Lung Cancer", and "Diabetes" (Fig. 3E), whereas "Sick", "Lung Cancer", and "Diabetes" are the three search queries with highest-weights to decode "death" (Fig. 3F). Table 1 summarizes the rank for sum of weight. Measures for accuracy (SEM, rho and p value) among three types of selections of predictors, including situations with all 19 predictors, 10 periodic predictors, and the 10 highest-weighted predictors, as shown in Fig. 2, are summarized and compared in Table 2.

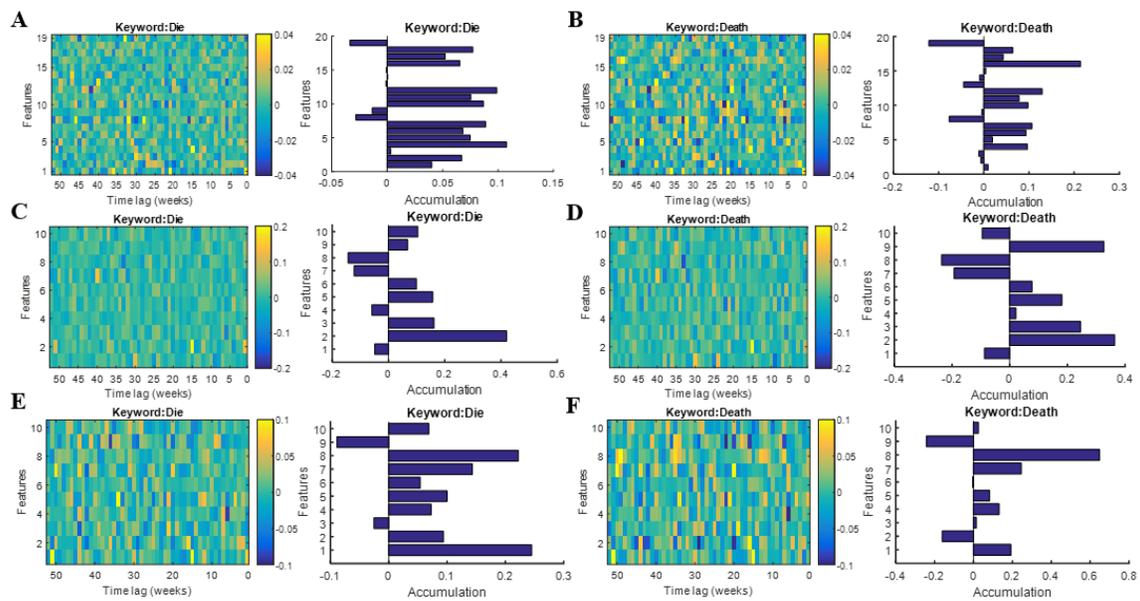

**Fig. 3.** Features importance for mortality forecasting. Three different types of selections of search terms are applied. The weight distribution across features of "die" and "death" and sum of weight for features using all 19 predictors are presented in (A) and (B), respectively. (C) and (D) show the features importance using only the top ten most periodic predictors. (E) and (F) demonstrate the feature performance using only the ten predictors with highest weighting in (A) and (B).

**Table 2.** Comparison of measures for accuracy (MSE, rho and p value) among three types of selections of predictors, including situations with all 19 predictors, 10 periodic predictors, and the 10 highest-weighted predictors.

| Combination | Keyword | MSE | rho | p value |
| --- | --- | --- | --- | --- |
| All 19 predictors | Die | 23.24 | 0.49 | < 0.001 |
| | Death | 52.61 | 0.25 | < 0.001 |
| 10 periodic predictors | Die | 31.42 | 0.44 | < 0.001 |
| | Death | 78.00 | 0.07 | 0.32 |
| 10 highest-weighted predictors | Die | 25.50 | 0.48 | < 0.001 |
| | Death | 47.93 | 0.31 | < 0.001 |

## 4. Conclusion

In this work, we present the key search queries from Google Trends can be used to forecast mortality related terminologies, such as "die" and "death". The effects of periodic patterns and weight distributions of each search queries are investigated and compared. This study provides guidance for developing a mortality surveillance system in the future.

**Acknowledgments**

This work was supported by Taiwan Ministry of Science and Technology (NSC 101-2221-E-157-007-MY2).